\documentclass[a4paper,11pt]{article}

\usepackage{amsmath,amssymb,color,graphics,epsfig,cite}
\usepackage{titlesec}
\usepackage[titletoc,title]{appendix}

\usepackage{amsfonts}

\newcommand{\be}{\begin{equation}}
\newcommand{\ee}{\end{equation}}
\newcommand{\bea}{\setlength\arraycolsep{2pt} \begin{eqnarray}}
\newcommand{\eea}{\end{eqnarray}}

\def\fft#1#2{{\frac{#1}{#2}}}

\def\0{{\sst{(0)}}}
\def\1{{\sst{(1)}}}
\def\2{{\sst{(2)}}}
\def\3{{\sst{(3)}}}
\def\4{{\sst{(4)}}}
\def\5{{\sst{(5)}}}
\def\6{{\sst{(6)}}}
\def\7{{\sst{(7)}}}
\def\8{{\sst{(8)}}}
\def\sst#1{{\scriptscriptstyle #1}}

\begin{document}

\begin{center}
{\Large {\bf The Upper Bound of Event Horizon Formation Time in Generalized Oppenheimer--Snyder Collapse}}

\vspace{20pt}

Zhi-Chao Li, H.~Khodabakhshi and H.~L\"u

\vspace{10pt}

{\it Center for Joint Quantum Studies, Department of Physics,\\
School of Science, Tianjin University, Tianjin 300350, China }

\vspace{40pt}

\underline{ABSTRACT}
\end{center}

We prove that, in the framework of the Oppenheimer-Snyder collapse, the Schwarzschild exterior maximizes the event horizon formation time $\Delta T_{\text{eh}}=\frac{19}{6}m$ among all asymptotically flat, static, spherically-symmetric black holes with the same ADM mass $m$ that satisfy the weak energy condition. This bound extends the typical black hole inequalities--such as the Penrose inequality, which constrains spatial geometry--to temporal setting.

\vfill{\footnotesize lizc@tju.edu.cn \ \ \  h\_khodabakhshi@tju.edu.cn
\ \ \ mrhonglu@gmail.com}

\thispagestyle{empty}
\pagebreak

\section{Introduction}

Gravitational collapse provides the theoretical foundation for black hole formation, a central prediction of general relativity~\cite{WaldGR,MTW,HawkingEllis}. The classic Oppenheimer-Snyder (OS) model~\cite{8} gives the first fully relativistic description of a pressureless, homogeneous star collapsing into a Schwarzschild black hole. In this scenario, the interior is modeled by a Friedmann--Lema\^{\i}tre--Robertson--Walker (FLRW) spacetime, while the exterior is vacuum Schwarzschild~\cite{Schwarzschild1916}; the two are matched smoothly across the stellar surface via the Darmois--Israel junction conditions~\cite{Darmois,Israel:1966rt,PoissonToolkit}. Throughout, we focus on collapse scenarios consistent with standard energy conditions~\cite{HawkingEllis,WaldGR, PoissonToolkit} and leading to black hole formation (i.e., without naked singularities), assuming the weak cosmic censorship conjecture~\cite{Penrose1969}.

For the FLRW interior spacetime, we will show that OS-type consistent matching to a static, spherically symmetric exterior implies the constraint $g_{tt}g_{rr}=-1$. By the ``generalized'' OS collapse, one goes beyond the vacuum Schwarzschild exterior. The OS-type matching and related collapse models have been explored for charged exteriors such as the Reissner--Nordstr\"om black hole~\cite{Raychaudhuri:1975efc} and for spacetimes with a cosmological constant (Schwarzschild--(A)dS/Kottler)~\cite{Kottler1918, MarkovicShapiroLambda}, among other settings. These generalizations examine how external fields or effective matter profiles influence horizon formation and the role of energy conditions; see, e.g.,~\cite{PoissonToolkit,HawkingEllis}. More recently, OS-type collapse into regular black holes has also been investigated in asymptotic-safety/scale-dependent frameworks; see, e.g.,~\cite{Bonanno:2024PRL,Harada:2025PRD,Shojai:2022CQG, Hassannejad:2025PRD}.

In \cite{OurPRD}, we studied OS collapse for a wide class of static, spherically symmetric black hole exteriors in a general setting, including Reissner--Nordstr\"om and Schwarzschild--(A)dS examples, improving upon earlier  work~\cite{Raychaudhuri:1975efc, MarkovicShapiroLambda} by clarifying the conditions for physically consistent collapse and systematically computing horizon formation and related properties. We found that the collapse must start from a radius no smaller than a certain minimum value, $R_{0,\text{min}}$; otherwise, the event horizon would have already formed before the collapse begins, which is incompatible with a horizon-free initial slice. We also showed that the proper time it takes for the event horizon to form on the stellar surface (denoted $\Delta T_{\text{eh}}$) depends only on the exterior geometry and not on the chosen starting radius $R_0$. Motivated by explicit examples, together with the fact that the Schwarzschild geometry has the largest horizon size at fixed ADM mass~\cite{Lu:2019zxb,Yang:2019zcn}, we made the following conjecture:
\begin{quote}
\textit{Within the framework of OS collapse, for all asymptotically flat, static, spherically symmetric black hole spacetimes satisfying the weak energy condition and sharing the same ADM mass~\cite{ADM}, the Schwarzschild geometry maximizes both $R_{0,\text{min}}$ and $\Delta T_{\text{eh}}$.}
\end{quote}
This conjecture suggests an additional extremal property of the Schwarzschild geometry.

In this paper, we provide a proof of the conjecture. By expressing the collapse dynamics in terms of the Misner--Sharp mass $M(R)$~\cite{MisnerSharp} and introducing dimensionless variables, we reformulate the conditions defining $R_{0,\text{min}}$ and $\Delta T_{\text{eh}}$ as constraints on a probability measure derived from the exterior geometry. In four-dimensional Einstein gravity, for a static, spherically symmetric exterior with diagonal stress tensor, the weak energy condition implies in particular a nonnegative exterior energy density $\rho(R)\ge 0$~\cite{HawkingEllis,WaldGR,PoissonToolkit}. This is the key input in our argument: it guarantees that the dimensionless mass function $\mu=M(R)/m$ is nondecreasing and bounded above by unity. We perform a stochastic comparison~\cite{ShakedShanthikumar,MullerStoyan} between the general collapse measure $\nu$ and its Schwarzschild counterpart, from which we deduce that the expectation value $\langle x\rangle_\nu=\Delta T_{\text{eh}}/(2m)$ is maximized precisely in the Schwarzschild case.

The paper is organized as follows. In Section~\ref{sec:OS}, we review the generalized collapse setup~\cite{OurPRD} and derive the equations governing the stellar surface and the event horizon using standard matching methods~\cite{Darmois,Israel:1966rt,PoissonToolkit}. In Section~\ref{sec:bound}, we introduce a dimensionless formulation and prove the inequality by comparing probability measures via stochastic ordering~\cite{ShakedShanthikumar,MullerStoyan}, together with basic results from Lebesgue--Stieltjes integration~\cite{Billingsley,Folland}. We conclude in Section~\ref{sec:conclusion} with a discussion of physical implications and possible extensions.

\section{Oppenheimer--Snyder collapse}
\label{sec:OS}

We consider the generalized OS collapse of a homogeneous, pressureless star~\cite{OurPRD}. The interior is described by a spatially flat FLRW metric
\begin{equation}\label{frw}
ds^2=-d\tau^2+a(\tau)^2\bigl(dr_c^2+r_c^2 d\Omega^2\bigr),
\end{equation}
where $\tau$ is the proper time of comoving observers, $r_c$ is the comoving radial coordinate, and $a(\tau)$ is the scale factor. Introducing the physical radius $r=a(\tau)r_c$, this metric takes the Painlev\'e--Gullstrand (PG) form~\cite{Painleve1921,Gullstrand1922}
\begin{equation}\label{frwpg}
ds^2=-d\tau^2+\bigl(dr-rH(\tau)\,d\tau\bigr)^2+r^2 d\Omega^2,
\end{equation}
with Hubble parameter $H=\dot a/a=\dot r/r$.

The exterior spacetime is described by the static, spherically symmetric metric
\begin{equation}\label{ext_metric}
ds^2=-h(r)\,dt^2+\frac{dr^2}{f(r)}+r^2 d\Omega^2.
\end{equation}
Let the stellar surface be the timelike hypersurface $\Sigma:\ r=R(\tau)$, parametrized by the proper
time $\tau$ of comoving observers in the homogeneous FLRW interior.  The Darmois--Israel junction conditions require continuity of the extrinsic
curvature across $\Sigma$~\cite{Darmois,Israel:1966rt,PoissonToolkit}. A direct computation for the exterior
gives the nontrivial components (prime denotes $d/dr$ evaluated at $r=R(\tau)$)
\begin{equation}\label{eq:K_components_ext}
K^{\theta}{}_{\theta}=K^{\phi}{}_{\phi}=\frac{\sqrt{f+\dot R^{\,2}}}{R},\qquad
K^{\tau}{}_{\tau}=\frac{1}{2\sqrt{f+\dot R^{\,2}}}
\left(2\ddot R+(f+\dot R^{\,2})\frac{h'}{h}-\dot R^{\,2}\frac{f'}{f}\right).
\end{equation}
For a spatially flat homogeneous FLRW interior with a comoving boundary, one has
$K^{\theta}{}_{\theta}=K^{\phi}{}_{\phi}=1/R$ and $K^{\tau}_{\tau}=0$ on $\Sigma$. Matching the angular components therefore yields
\begin{equation}\label{eq:Rdot_marginal}
\sqrt{f(R)+\dot R^{\,2}}=1
\qquad\Longrightarrow\qquad
\dot R^{\,2}=1-f(R),
\end{equation}
and differentiating \eqref{eq:Rdot_marginal} gives $\ddot R=-\tfrac12 f'(R)$. Substituting these relations
into the matching condition $K^{\tau}{}_{\tau}=0$ in \eqref{eq:K_components_ext}, implies
\[
\frac{h'}{h}=\frac{f'}{f}\qquad\Longrightarrow\qquad h(r)=C\,f(r),
\]
with constant $C$. For asymptotically flat exteriors, the standard time normalization $h(\infty)=1$ together
with $f(\infty)=1$ fixes $C=1$, and hence
\begin{equation}\label{hf_equal}
h(r)=f(r).
\end{equation}
Let $r=r_+$ denote the outer horizon defined by $f(r_+)=0$. By generalized OS collapse, we do not assume vacuum outside the star. In this paper, we focus on $f(r)$ that correspond to any asymptotically flat black hole geometry (e.g.\ Reissner--Nordstr\"om) that satisfies the weak energy condition.

Introducing the infalling Painlev\'e--Gullstrand (PG) time coordinate $T$, the exterior metric with
$h=f$ in~\eqref{hf_equal} can be cast into the standard form
\begin{equation}\label{metpg}
ds^2=-dT^2+\bigl(dr+\sqrt{1-f(r)}\,dT\bigr)^2+r^2 d\Omega^2.
\end{equation}
Along the stellar surface $r=R(\tau)$ we identify $T=\tau$. Choosing the collapsing branch of
\eqref{eq:Rdot_marginal}, we have
\begin{equation}\label{req}
\dot R=-\sqrt{1-f(R)},
\end{equation}
where $\dot R\equiv dR/d\tau$. We assume the collapse begins at $\tau=0$ (equivalently $T=0$ on the
surface) from an initial radius $R_0$, so $R(0)=R_0$. Integrating~\eqref{req} gives
\begin{equation}\label{tsr}
T(R)=-\int_{R_0}^{R}\frac{dR'}{\sqrt{1-f(R')}}.
\end{equation}
The interior event-horizon radius $R_{\text{eh}}(R)$ is obtained by tracking outgoing radial null geodesics backward from the exterior horizon, with boundary condition $R_{\text{eh}}(R_+)=R_+$.
As shown in~\cite{OurPRD}, this yields
\begin{equation}\label{rreh}
R_{\text{eh}}(R)
=R\left(1-\int_{R_+}^{R}\frac{dR'}{R'\sqrt{1-f(R')}}\right).
\end{equation}
The event horizon is born at the center when $R_{\text{eh}}=0$. This defines the \emph{minimum initial
radius} $R_{0,\text{min}}$, implicitly given by
\begin{equation}\label{r0}
1=\int_{R_+}^{R_{0,\text{min}}}\frac{dR}{R\sqrt{1-f(R)}}.
\end{equation}
If $R_0<R_{0,\text{min}}$, the event horizon would already exist at $\tau=0$, violating the assumption of
a horizon-free initial slice.

The \emph{event horizon formation time}, measured in proper time by an observer on the stellar surface,
is the interval between horizon birth at the center and the surface crossing $R_+$:
\begin{equation}\label{ttt}
\Delta T_{\text{eh}}=\int_{R_+}^{R_{0,\text{min}}}\frac{dR}{\sqrt{1-f(R)}}.
\end{equation}
This quantity depends only on the exterior geometry, not on $R_0$, making it a natural timescale for
collapse. For the Schwarzschild black hole, one finds
\begin{equation}\label{schrt}
R_{0,\text{min}}^{\text{(sch)}}=\frac{9}{2}m\,, \qquad
\Delta T_{\text{eh}}^{\text{(sch)}}=\frac{19}{6}m\,.
\end{equation}
In~\cite{OurPRD}, we conjectured that, under a suitable energy condition, among all asymptotically flat, static, spherically-symmetric black holes with the same ADM mass $m$ that admit OS-type collapse, the Schwarzschild solution maximizes both $R_{0,\text{min}}$ and $\Delta T_{\text{eh}}$. We identify the relevant condition and prove the conjecture in the next section.

\section{Upper bound on the event horizon formation time}
\label{sec:bound}

\subsection{The setup}

We work in four-dimensional Einstein gravity (with $G=c=1$)~\cite{WaldGR,MTW} and consider a static, spherically symmetric, asymptotically flat exterior whose metric functions $h=f$ can be written in the Misner--Sharp form $f(R)=1-\frac{2M(R)}{R}$, where
\begin{equation}\label{eq:MS_mass_def}
M(R)=\frac{R}{2}\bigl(1-f(R)\bigr)
\end{equation}
is the Misner--Sharp mass~\cite{MisnerSharp}. For a static, spherically-symmetric matter source with energy--momentum tensor $T^{a}{}_{b}=\mathrm{diag}\bigl(-\rho,\, p_R,\, p_T,\,p_T \bigr)$, the Einstein equations imply~\cite{WaldGR,PoissonToolkit}
\begin{equation}\label{eq:Mprime_rho}
\fft{d}{dR} M(R)=4\pi R^{2}\rho(R).
\end{equation}
Hence, under the weak energy condition so that $(\rho(R) \geq 0)$,  $M(R)$ is a monotonically nondecreasing function of $R$. For asymptotically-flat spacetimes, the Misner--Sharp mass satisfies~\cite{ADM,MisnerSharp,PoissonToolkit}
\begin{equation}\label{eq:MS_boundary}
M(R_+)=\frac{R_+}{2},\qquad \lim_{R\to\infty} M(R)=m,
\end{equation}
where $m$ is the ADM mass and $R_+$ is the (outer) horizon radius defined by $f(R_+)=0$. Thus, under the weak energy condition, we have
\begin{equation}\label{eq:M_bounds}
\frac{R_+}{2}\le M(R)\le m,  \qquad \hbox{for}\qquad R\ge R_+.
\end{equation}
To expose the underlying structure, we introduce the dimensionless radial variables
\begin{equation}\label{eq:dimless_defs}
x:=\frac{R}{2m},
\qquad
x_+:=\frac{R_+}{2m},
\qquad
x_0:=\frac{R_{0,\text{min}}}{2m},
\end{equation}
and the dimensionless mass function
\begin{equation}\label{eq:mu_def}
\mu(x):=\frac{M(R)}{m},
\qquad (R=2mx).
\end{equation}
It is easy to verify that the inequality \eqref{eq:Mprime_rho} from the weak energy condition also implies that $\mu(x)$ is nondecreasing.  It follows that
\begin{equation}\label{eq:xplus_le1}
0<x_+=\mu(x_+)\le \mu(x) \le \mu(\infty) =1,\qquad \hbox{for}\qquad x\ge x_+.
\end{equation}
In terms of $(x,\mu)$, the equations \eqref{r0} and \eqref{ttt} become
\be
1 =\int_{x_+}^{x_0}\frac{dx}{\sqrt{x\, \mu(x)}}\,,\qquad \Delta T_{\text{eh}}
=2m\int_{x_+}^{x_0}\sqrt{\frac{x}{\mu(x)}}\,dx. \label{eq:Teh_mu}
\ee

\subsection{Proof of $R_{0,\text{min}}\le \frac{9}{2}m$}

It follows from \eqref{eq:xplus_le1} that the first equation in \eqref{eq:Teh_mu} becomes an inequality
\begin{equation}\label{eq:x0_bound_step}
1=\int_{x_+}^{x_0}\frac{dx}{\sqrt{x\,\mu(x)}}
\ge \int_{x_+}^{x_0}\frac{dx}{\sqrt{x}}
=2\bigl(\sqrt{x_0}-\sqrt{x_+}\bigr)\ge 2\bigl(\sqrt{x_0}-1\bigr).
\end{equation}
In the above, the two inequalities follow from $\mu(x)\le 1$ and $x_+\le 1$ respectively.
This immediately implies
\begin{equation}\label{eq:x0_max}
\sqrt{x_0}\le \frac{3}{2},
\qquad\text{or equivalently}\qquad
x_0\le \frac{9}{4}.
\end{equation}
The inequality is saturated by the Schwarzschild black hole. Note that the result implies
\begin{equation}\label{eq:interval_containment}
x\in[x_+,x_0]\subset(0,9/4].
\end{equation}

\subsection{Proof of $\Delta T_{\text{eh}} \le \frac{19}{6}m$}

To prove this statement, it is useful to interpret the left equation of \eqref{eq:Teh_mu} as defining a normalized measure. To see this, we define a measure $\nu$ on $\mathbb{R}$ by
\begin{equation}\label{eq:nu_def}
d\nu(x):=\frac{dx}{\sqrt{x\,\mu(x)}}, \qquad \text{for }\qquad x\in[x_+,x_0],
\end{equation}
and set $d\nu(x)=0$ outside $[x_+,x_0]$. Then the left equation of \eqref{eq:Teh_mu} is precisely the normalization
\begin{equation}\label{eq:nu_normalized}
\int_{\mathbb{R}} d\nu=\int_{x_+}^{x_0} d\nu(x)=1,
\end{equation}
so that $\nu$ can be viewed as a probability measure. The event horizon formation time, given by the right equation of \eqref{eq:Teh_mu}, can now be written as
\begin{equation}\label{eq:Teh_expectation}
\frac{\Delta T_{\text{eh}}}{2m}
=\int_{x_+}^{x_0}\sqrt{\frac{x}{\mu(x)}}\,dx
=\int_{x_+}^{x_0} x\,d\nu(x) = \langle x\rangle_\nu.
\end{equation}
In other words, the quantity $\Delta T_{\text{eh}}/(2m)$ is an expectation value of the dimensionless radial coordinate, or the first moment of $x$ under $\nu$~\cite{ShakedShanthikumar,MullerStoyan}. 

To proceed, we define two cumulative distribution functions (CDFs)
\[
F(x)=\int_{-\infty}^x d\nu,
\qquad
F_{\text{sch}}(x)=\int_{-\infty}^x d\nu_{\text{sch}},
\]
Explicitly,
\begin{equation}\label{eq:CDF_general}
F(x)=
\begin{cases}
0, & x<x_+,\\[4pt]
\displaystyle \int_{x_+}^{x}\frac{dx'}{\sqrt{x'\,\mu(x')}}, & x_+\le x\le x_0,\\[8pt]
1, & x\ge x_0,
\end{cases}
\end{equation}
and
\begin{equation}\label{eq:CDF_sch}
F_{\text{sch}}(x)=
\begin{cases}
0, & x<1,\\[4pt]
\displaystyle \int_{1}^{x}\frac{dx'}{\sqrt{x'}}=2(\sqrt{x}-1), & 1\le x\le \dfrac{9}{4},\\[8pt]
1, & x\ge \dfrac{9}{4}.
\end{cases}
\end{equation}
It is clear both $F$'s are continuous functions of variable $x$. Furthermore, we claim that
\begin{equation}\label{eq:stochastic_order}
F(x)\ge F_{\text{sch}}(x)
\qquad \text{for all } x\in\mathbb{R}.
\end{equation}
Indeed: (i) if $x<x_+$, then $F(x)=0$ and also $F_{\text{sch}}(x)=0$ because $x_+\le 1$;
(ii) if $x_+\le x<1$, then $F(x)>0$ while $F_{\text{sch}}(x)=0$;
(iii) if $1\le x\le x_0$, then
\[
F(x)-F_{\text{sch}}(x)
=\int_{x_+}^{1}\frac{dx'}{\sqrt{x'}\,\sqrt{\mu(x')}}
+\int_{1}^{x}\left(\frac{1}{\sqrt{x'}\,\sqrt{\mu(x')}}-\frac{1}{\sqrt{x'}}\right)dx'\ge 0,
\]
since $\mu(x')\le 1$;
(iv) if $x_0<x<9/4$, then $F(x)=1$ while $F_{\text{sch}}(x)<1$;
(v) if $x\ge 9/4$, both equal $1$.
Thus \eqref{eq:stochastic_order} holds and expresses the first-order stochastic dominance~\cite{ShakedShanthikumar,MullerStoyan}.

To compare first moments, note that $\nu$ is supported in $[x_+,x_0]$ and $\nu_{\text{sch}}$ in $[1,9/4]$. Since $x_0\le 9/4$, for any integrable $\varphi$ we have
\begin{equation}\label{eq:support_reduction}
\int_{0}^{9/4}\varphi(x)\,d\nu(x)=\int_{x_+}^{x_0}\varphi(x)\,d\nu(x),
\qquad
\int_{0}^{9/4}\varphi(x)\,d\nu_{\text{sch}}(x)=\int_{1}^{9/4}\varphi(x)\,d\nu_{\text{sch}}(x).
\end{equation}
Next, we use the Lebesgue--Stieltjes integration-by-parts identity~\cite{Billingsley,Folland}: if $\nu$ is a probability measure supported in $[0,X]$ with CDF $F(x)$, then
\begin{equation}\label{eq:moment_identity}
\int_{0}^{X} x\,d\nu(x)
=\Bigl[xF(x)\Bigr]_{0}^{X}-\int_{0}^{X}F(x)\,dx
=X-\int_{0}^{X}F(x)\,dx.
\end{equation}
Applying \eqref{eq:moment_identity} to $\nu$ and $\nu_{\text{sch}}$ with $X =9/4$, and using $F(0)=F_{\text{sch}}(0)=0$ and $F(9/4)=F_{\text{sch}}(9/4)=1$, we obtain
\begin{equation}\label{eq:moment_identity_applied}
\int_{0}^{9/4} x\,d\nu(x)=\frac{9}{4}-\int_{0}^{9/4}F(x)\,dx,
\qquad
\int_{0}^{9/4} x\,d\nu_{\text{sch}}(x)=\frac{9}{4}-\int_{0}^{9/4}F_{\text{sch}}(x)\,dx.
\end{equation}
Since $F(x)\ge F_{\text{sch}}(x)$ on $[0,9/4]$ by \eqref{eq:stochastic_order}, integrating and subtracting from $9/4$ yields
\begin{equation}\label{eq:moment_inequality}
\int_{0}^{9/4} x\,d\nu(x)\le \int_{0}^{9/4} x\,d\nu_{\text{sch}}(x) = \int_{1}^{9/4} x\,\frac{dx}{\sqrt{x}} =\frac{19}{12}.
\end{equation}
Combining \eqref{eq:moment_inequality} with \eqref{eq:Teh_expectation} and \eqref{eq:support_reduction} gives
\[
\frac{\Delta T_{\text{eh}}}{2m}
=\int_{x_+}^{x_0} x\,d\nu(x)
=\int_{0}^{9/4} x\,d\nu(x)
\le \frac{19}{12},
\qquad\text{i.e.}\qquad
\Delta T_{\text{eh}}\le \frac{19}{6}m.
\]
The saturation occurs when the exterior geometry is exactly Schwarzschild.

\section{Conclusion}
\label{sec:conclusion}

In this work, we have proven the conjecture proposed in our earlier paper~\cite{OurPRD}. Within the generalized OS collapse framework~\cite{8}, we assume a static, spherically symmetric, asymptotically flat exterior in four-dimensional Einstein gravity~\cite{WaldGR,MTW} and impose nonnegative exterior energy density $\rho(R)\ge 0$, as implied by the weak energy condition in the present setting. Furthermore, the consistency of the OS collapse requires that $h=f$ for the metric ansatz \eqref{ext_metric}. Under these assumptions, the Schwarzschild geometry maximizes both the minimum admissible initial radius $R_{0,\text{min}}$ (defined by the requirement of a horizon-free initial slice) and the event-horizon formation time $\Delta T_{\text{eh}}$ among all exteriors with the same ADM mass $m$, i.e.
\begin{equation}
\Delta T_{\text{eh}} \le \frac{19}{6}m,
\end{equation}
Equivalently, within this class of models, Schwarzschild realizes the \emph{slowest} classical formation of an event horizon.

Many bounds in black hole physics have been proposed in literature. The most famous one is the Penrose inequality \cite{Bray:2003ns,Mars:2009cj}. (See also \cite{Dain:2011pi,Lu:2019zxb,Yang:2019zcn,Khodabakhshi:2022jot,Lu:2025iky}). These bounds typically restrict the spatial size of black hole by the conserved quantities of a black hole. Our inequality extends these traditional spatial bounds to involve time. It is of great interest to investigate whether the bound derived from the simplified OS collapse applies to more general realistic black hole collapses. The factor $19/6\sim 3.167$ is also intriguing since it gives slightly longer time than the longest time $\pi m$ that an object can stay inside the Schwarzschild black hole before falling into the singularity. Finally, the probabilistic framework introduced in this paper may provide a useful template for studying horizon formation times in other collapse models and in more general gravitational theories.

\section*{Acknowledgement}

This work is supported in part by the National Natural Science Foundation of China (NSFC) grants No.~W2533015 No.~12375052 and No.~11935009, as well as by the Tianjin University Self-Innovation Fund Extreme Basic Research Project Grant No.~2025XJ21-0007.

\end{document}